\documentclass[fleqn,twoside]{article}
\usepackage[headings]{espcrc2}

\usepackage{graphicx}
\usepackage[figuresright]{rotating}

\newcommand{\AmS}{{\protect\the\textfont2
  A\kern-.1667em\lower.5ex\hbox{M}\kern-.125emS}}
\def\be{\begin{eqnarray}}
\def\ee{\end{eqnarray}}
\def\beq{\begin{eqnarray}}
\def\eeq{\end{eqnarray}}

\newcommand{\ice}[1]{\relax}

\newcommand{\GeV}{{\rm GeV}}

\newcommand{\nn}{\nonumber}

\title{Sum rules for $B^0 - \bar B^0$ mixing at NLO of pQCD}

\author{A.~A.~Pivovarov\address[INR RAS]{Institute for Nuclear Research of the Russian
Academy of Sciences, 117312 Moscow, Russia}%
        \thanks{Talk given at 11th International QCD 04 Conference, 
Montpellier 5-10th July 2004; supported in part by 
the INTAS grant under contract 03-51-4007
and Volkswagen Foundation grant \#~I/77 788.} 
}

\runtitle{Sum rules for $B^0 - \bar B^0$ mixing}
\runauthor{A.~A.~Pivovarov}

\begin{document}

\begin{abstract}
The matrix element of $B^0 -\bar B^0$ mixing 
is evaluated from QCD sum rules for three-point correlators
with next-to-leading order
accuracy in pQCD~\cite{main1}.
\vspace{1pc}
\end{abstract}

\maketitle

\begin{itemize}
\item Introduction
\item Phenomenology of $B^0 -\bar B^0$ mixing 
\item Hadronic $B_B$ parameter
\item pQCD analysis at three loops
\item Results for $B_B$
\item Conclusion
\end{itemize}

\section{Introduction}
Particle-antiparticle mixing in systems of 
neutral mesons of different flavors 
\[
sd:~K^0 - \bar K^0;\quad cu:~D^0 - \bar D^0;\quad bd:~B^0 - \bar B^0
\]
is the primary source of CP violation studies. 
Historically, the study of $K^0 - \bar K^0$ mixing was very
fruitful for elementary particle physics. It provided deep insights
into delicate questions of weak interactions and gave first convincing
proof of possibility for CP violation.
The thorough quantitative analysis of mixing in the kaon system
strongly constrained the physics of heavy particles and possible
scenarios for the extension of the light quark sector of the theory.
The numerical value of the mass splitting between the eigenstates 
of the effective Hamiltonian for the kaon pair has been used 
to estimate the numerical value of the charm quark mass
from the requirement of GIM cancellation before the 
experimental discovery of charm~(see e.g.~\cite{kkreviewOLD}). 
The proof of existing CP violation in the interaction of particles 
happens to be very important for our understanding of the structure of
Universe and its evolution. For example, the violation of CP symmetry
is one of necessary conditions for generating the baryon asymmetry of the
Universe~\cite{Sakharov:1967dj}.

Presently the experimental studies of phenomena of CP violation and mixing 
are active in the area of heavy mesons for which they are considered more 
promising. The charmed mesons $D (\bar u c)$ have recently 
attracted some interest and provided encouraging results~\cite{Petrov:2002is}. 
The systems of $B_d (\bar d b)$ and $B_s (\bar s b)$ 
mesons are the current laboratory for performing 
a precision analysis of CP violation and mixing both experimentally and 
theoretically~\cite{buhalla,reviewBB}. 
The experimental study is now under way in dedicated experiments 
at SLAC (BABAR collaboration, e.g.~\cite{Aubert:2002rg,Aubert:2002ic})
and 
KEK~(BELLE collaboration e.g.~\cite{Abe:2000yh,Abashian:2001pa}).

\section{Phenomenology of $B^0 -\bar B^0$ mixing} 
Mixing in a system of neutral pseudoscalar mesons 
is phenomenologically described by a 2x2 effective Hamiltonian 
$H_{eff}=\left(M-i\Gamma/2\right)_{ij}$, $\{i,j\}=\{1,2\}$ where $M$ is
related to the mass spectrum of the system and $\Gamma$ describes the
widths of the mesons. 
The time evolution of the system state vector $(B^0,\bar B^0)$
is governed by the equation
\[
i\frac{d}{dt}\left(
\begin{array}{c}
B^0\\
\bar B^0
\end{array}
\right)=H_{eff}
\left(
\begin{array}{c}
B^0\\
\bar B^0
\end{array}
\right) .
\]
The $\Delta B=2$ flavor violating interactions
generate the non-diagonal terms in the effective Hamiltonian. 
The mass difference
$\Delta m = M_{heavy}-M_{light}\approx 2\left|M_{12}\right|$ has been
precisely measured
$\Delta m= 0.489\pm0.005(stat)\pm 0.007(syst)~ps^{-1}$~\cite{PDG}.
This is an important 
observable which can be used to extract the top quark 
CKM parameters provided that the theoretical formulae are
sufficiently accurate.
The main difficulty of the analysis is to account for the effects of strong
interactions.

The effective $\Delta B = 2$ Hamiltonian is known at
next-to-leading order (NLO) in QCD perturbation theory of the Standard
Model~\cite{Buras}
\begin{eqnarray}
\label{hamilt}
&&H_{\mbox{eff}}^{\triangle B = 2} = \frac{G_F^2M_W^2}{4 \pi^2}
\left({V_{tb}}^{*}V_{td}\right)^2 \eta_B S_0(x_t)\\ \nn
&&\times\left[\alpha_s^{(5)}(\mu)\right]^{-6/23} 
\left[1+\frac{\alpha_s^{(5)}(\mu)}{4 \pi} J_5 \right] {\cal O}(\mu) .
\nonumber
\end{eqnarray}
Here $G_F$ is a Fermi constant, $M_W$ is the $W$-boson mass, 
$\eta_B=0.55\pm0.1$~\cite{Buras:1990fn}, $J_5=1.627$ in the
naive dimensional regularization (NDR) 
scheme, $S_0(x_t)$ is the Inami-Lim function~\cite{Inami:1980fz}, and
${\cal O}(\mu)=(\bar b_L\gamma_{\sigma}d_L)(\bar b_L\gamma_{\sigma}d_L)(\mu)$
is a local four-quark operator at the normalization point~$\mu$. 
Mass splitting of heavy and light mass eigenstates is
\begin{eqnarray}
\label{offdiag}
\Delta m =2 
|\langle\bar B^0|H_{\mbox{eff}}^{\triangle B = 2}|B^0 \rangle| .
\end{eqnarray} 
The largest uncertainty in the calculation of the mass splitting  
is introduced by the hadronic matrix element
${\cal A} = \langle\bar B^0|{\cal O}(\mu)|B^0\rangle$ 
that is poorly known~\cite{PDG}. 

\section{Hadronic $B_B$ parameter}
Theoretical evaluation of the hadronic 
matrix element ${\cal A}$ is a genuine non-perturbative
task which should 
be approached with some non-direct techniques. The simplest approach 
(``factorization''~\cite{Gaillard:1974hs}) 
reduces the matrix element ${\cal A}$ to the product of simpler matrix
elements measured in leptonic $B$ decays
\[
{\cal A}^{f} = \frac{8}{3}
\langle\bar B^0|\bar b_L\gamma_{\sigma}d_L|0\rangle
\langle 0|\bar b_L \gamma^{\sigma}d_L|B^0\rangle 
= \frac{2}{3} f_B^2 m_B^2
\]
where the leptonic decay constant $f_B$ is defined by
the relation 
$\langle 0|\bar b_L \gamma_\mu d_L|B^0({\bf p})\rangle 
= i p_\mu f_{B}/2$ 
and $m_B$ is the $B^0$ meson mass.
A deviation from the factorization ansatz is usually described by the parameter
$B_B$ defined as
$
{\cal A} = B_B {\cal A}^{f}
$;
in factorization $B_B=1$.
The evaluation of this parameter (and the
analogous parameter $B_K$ of $K^0 - \bar K^0$ mixing) has long
history. Many different results were obtained within approaches based
on quark models, unitarity, ChPT. The approach of direct numerical
evaluation on the lattice has also been used.
The corresponding results can be found in the 
literature~\cite{Ba,ope-three-kk,bb-three,Reind,Nar,Mhov,Pi90re,lattice,Hiorth}.

In my talk I report on the results 
of the calculation of the hadronic mixing matrix elements 
using Operator Product Expansion (OPE) and QCD 
sum rule techniques for three-point 
functions~\cite{ope-three-kk,bb-three,kkaplhas,svz,fesr}. 
This approach is very close in spirit to lattice computations~\cite{lattice}.
The advantages of this approach are: 
\begin{itemize}
\item It is a model-independent, first-principles method. 
The difference with the lattice approach is that the QCD sum rule
approach uses an asymptotic expansions of a Green's function
computed analytically 
while on the lattice the function itself can be numerically computed
provided the accuracy of the technique is sufficient. 
\item 
The sum rule techniques provide a consistent way of taking 
into account perturbative corrections
which is needed to restore the RG invariance of
physical observables usually violated in the factorization
approximation~\cite{kkaplhas}.
\end{itemize}
A concrete realization of the sum rule method used in the analysis
consists in 
the calculation of the moments of the 
three-point correlation function of the interpolating operators of 
the $B$-meson and the local operator ${\cal O}(\mu)$ responsible for 
the $B^0 - \bar B^0$ transitions. 

\section{pQCD analysis at three loops}
Consider the three-point correlation function
\begin{eqnarray}
\label{threepoint1}
&&\Pi(p_1,p_2)\\ \nn
&&=\int dx dy 
\langle 0|T J_{\bar B}(x) {\cal O}(0) \bar J_{B}(y)|0\rangle  
e^{i p_2 x - i p_1 y}
\end{eqnarray}
of the relevant $\Delta B=2$ operator ${\cal O}(\mu)$
and interpolating currents for the $B^0$-meson
$J_{B} = (m_b+m_d)\bar d i\gamma_5 b$. Here 
$m_b$ is the $b$ quark mass. The current $J_{B}$ is RG invariant and
$
J_{B} =\partial_\mu (\bar d \gamma_\mu\gamma_5 b)
$.
The main relevant property of this current is 
$
\langle 0|J_{B}(0)|B^0(p)\rangle =  f_{B} m_B^2
$
where $m_B$ is the $B$-meson mass.
A dispersive representation of the correlator reads
\be 
\label{dispdouble}
\Pi(p_1, p_2)=
\int\frac{\rho(s_1, s_2, q^2)ds_1 ds_2}{(s_1-p_1^2)(s_2-p_2^2)}
\ee
where $q=p_2-p_1$. For the analysis of $B^0 - \bar B^0$ mixing within
the sum rule framework this correlator 
can be computed at $q^2=0$. 

Phenomenologically the mixing matrix element
determines the contribution of the $B$-mesons in the form of a double
pole to the three-point correlator
\[
\label{phenrepr}
\Pi(p_1,p_2)\sim
\frac{\langle J_{\bar B}|\bar B^0\rangle}{m_{B}^2-p_1^2}
\langle\bar B^0|{\cal O}(\mu)|B^0\rangle
\frac{\langle B^0|\bar J_{B}\rangle}{m_{B}^2-p_2^2}.
\]
Because of technical difficulties of calculation, a practical way of 
extracting the $B^0 - \bar B^0$ matrix element
is to analyze the moments of the correlation function at 
${p_1^2=p_2^2=0}$ at the point $q^2=0$
\be 
\label{momentsdef}
M(i,j)
=\int \frac{\rho(s_1, s_2, 0)ds_1 ds_2}{s_1^{i+1} s_2^{j+1}}\, .\nn
\ee
A theoretical computation of these moments reduces to
an evaluation of single-scale vacuum diagrams 
and can be done analytically with available tools
for the automatic computation of multi-loop diagrams. 
Note that masses of light quarks are small 
(e.g.~\cite{Becchi:1980vz,Gasser:1982ap,lightmasses})
and can be accounted for as small perturbation. This makes it possible
to analyze the problem of mixing for $B_s$ mesons~\cite{narison2}.
\begin{figure}[t]
\begin{center}
\includegraphics[scale=0.5]{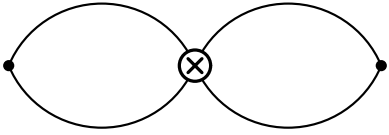}
\end{center}\caption{Perturbation theory diagram at LO}
\label{figLO}
\end{figure}
The leading contribution to the asymptotic expansion is given by 
the diagram shown in Fig.~\ref{figLO}. At the leading order in 
QCD perturbation theory the three-point function $\Pi(p_1, p_2)$
of Eq.~(\ref{threepoint1}) completely factorizes
\be
\Pi(p_1, p_2) = \frac{8}{3}\Pi_\mu (p_1)\Pi^\mu (p_2)
\ee
into a product of the two-point correlators $\Pi_\mu (p)$
\be 
\label{twopointscorr}
&& \Pi_\mu (p)=p_\mu \Pi(p^2)\\ \nn
&&=\int dx e^{i p x}
\langle 0|TJ_{\bar B}(x) \bar b_L \gamma_\mu d_L(0)|0\rangle .
\end{eqnarray}
At the LO the calculation of moments is straightforward since
the double spectral density is explicitly known
in this approximation. Indeed, using dispersion relation for the
two-point correlator 
\be 
&&\Pi(p^2) =\int_{m^2}^{\infty} \frac{\rho(s)ds}{s-p^2}, \\ \nn
&&
\rho(s)=\frac{3}{16\pi^2}m^2\left(1-\frac{m^2}{s}\right)^2
\ee
one obtains the LO double spectral density
in a factorized form
\[
\rho^{\rm LO}(s_1,s_2,q^2) =
\frac{8}{3} (p_1 \cdot p_2)\rho(s_1)\rho(s_2) .
\]
Thus, all PT contributions are of the factorizable form at the leading
order.
First non-factorizable contributions to Eq.~(\ref{dispdouble}) appear
at NLO. Of course, at NLO there are also the factorizable diagrams.
Note that the classification of diagrams in terms of their
factorizability is consistent as both classes
are independently gauge and RG invariant. 
\begin{figure}[t]
\begin{center}
\includegraphics[scale=0.4]{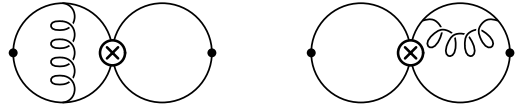}
\end{center}
\caption{Factorizable diagrams at NLO}
\label{figNLOfac}
\end{figure}
Consider first the NLO factorizable contributions 
that are given by the product of two-point 
correlation functions from Eq.~(\ref{twopointscorr}), 
as shown in Fig.~\ref{figNLOfac}.
Analytical expression for such contributions can be obtained as
follows. Writing 
$
\Pi(p^2)=\Pi_{\rm LO}(p^2)+\Pi_{\rm NLO}(p^2)
$
one finds
\[
\Pi_{\rm NLO}^f(p_1, p_2) = \frac{8}{3}(p_1.p_2)
(\Pi_{\rm LO}(p_1^2)\Pi_{\rm NLO}(p_2^2)
\]
\[
+\Pi_{\rm NLO}(p_1^2)\Pi_{\rm LO}(p_2^2)).
\]
The spectral density of the correlator
$\Pi_{\rm NLO}(p^2)$ is known analytically
that solves the problem of the NLO analysis in 
factorization. Even a NNLO analysis of factorizable 
diagrams is possible as several moments of two-point correlators are 
known analytically~\cite{chetmomnondiag}. 

\begin{figure}[t]
\begin{center}
\includegraphics[scale=0.5]{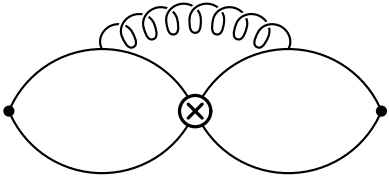}
\end{center}
\caption{A non-factorizable diagram at NLO}
\label{figNLOnonfac}
\end{figure}

The NLO analysis of non-factorizable contributions within perturbation
theory is the main point of 
my talk. The analysis amounts to the calculation of a set of 
three-loop diagrams (a typical diagram is presented in Fig.~\ref{figNLOnonfac}). 
These diagrams have been computed using the package MATAD for automatic calculation
of Feynman diagrams~\cite{matad}. The package is applicable only for
computation of scalar integrals.
the decomposition of the three-point amplitude
into scalars is known~\cite{Davyd}. 
The expression for the ``theoretical'' moments is
\be
\label{thfull}
M_{th}(i,j)=\frac{m^6 a_{ij}}{m^{2(i+j)}}
\left\{1+\frac{\alpha_s}{4\pi} \left(b^{f}_{ij}+b^{nf}_{ij}\right)\right\}
\ee
where the quantities $a_{ij}$, $b^{f}_{ij}$ and $b^{nf}_{ij}$
represent LO, NLO factorizable and NLO nonfactorizable contributions 
as shown in Figs.~\ref{figLO}-\ref{figNLOnonfac}. 
The NLO nonfactorizable contributions $b^{nf}_{ij}$ with 
$i+j\leq 7$ are analytically calculated in ref.~\cite{main1}
for the first time.
The calculation has been done with the system FORM~\cite{form}
and required about 24 hours of computing time on a dual-CPU 
2 GHz Intel Xeon machine. 
The analytical result for
the lowest finite moment $M_{th}(2,2)$ reads
\be
a_{22}=\frac{1}{(16\pi^2)^2}
\left(\frac{8}{3}\right), \quad
b^{f}_{22}=\frac{40}{3}+\frac{16\pi^2}{9}\, ,
\ee
\[
b^{nf}_{22}
\!=\!
S_2\frac{8366187}{17500}-\!\zeta_3\frac{84608}{875}
-\!\pi^2\frac{33197}{52500}
-\!\frac{426319}{315000}.
\]
Here
$
S_2=\frac{4}{9\sqrt{3}}{\rm Cl}_2
\left(\frac{\pi}{3}\right)=0.2604\ldots
$, $\zeta_3=\zeta(3)$, and $\mu^2=m^2$. 
Higher moments contain the same transcendental entries 
$S_2$, $\zeta_3$, $\pi^2$ with different
numerical coefficients. The numerical values 
for the moments are
$
b^{nf}_{ij}
$:
$
b^{nf}_{2(2345)}=\{0.68,1.22,1.44,1.56\}
$
and 
$
b^{nf}_{3(34)}=\{1.96,2.25\}
$.
The above theoretical results are used to 
extract the non-perturbative parameter $B_B$ from the sum rules 
analysis.

\section{Results for $B_B$}
\begin{figure}[t]
\begin{center}
\includegraphics[scale=0.4]{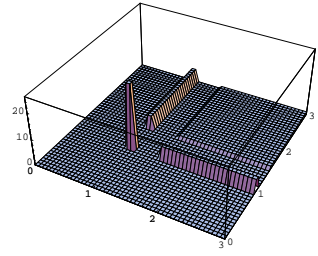}
\end{center}
\caption{A model of phenomenological spectrum}
\label{phenplot}
\end{figure}

The ``phenomenological'' side of the sum rules reads
\begin{eqnarray}
\label{phenfull}
M_{ph}(i,j)
= \frac{8}{3} B_B
\frac{f_B^4 m_B^2 }{m_B^{2(i+j)}}
+\ldots
\end{eqnarray}
where the contribution of the $B$-meson is displayed
explicitly. The remaining parts are the contributions due 
to higher resonances and 
the continuum which are suppressed due to the mass gap $\Delta$ in the 
spectrum model. A rough picture of the phenomenological spectrum is
given in Fig.~\ref{phenplot}.
For comparison we consider the factorizable approximation
for both ``theoretical''
\be
\label{thfact}
M_{th}^{f}(i,j)=\frac{m^6 a_{ij} }{m^{2(i+j)}}\left(1
+\frac{\alpha_s}{4\pi} b^{f}_{ij}\right)
\ee
and ``phenomenological'' moments, which, by construction,
are built from the moments of the two-point function
of Eq.~(\ref{twopointscorr})
\be
\label{phenfactprod}
M_{ph}^{f}(i,j)
= \frac{8}{3}\frac{f_B^4 m_B^2}{m_B^{2(i+j)}}+...
\end{eqnarray}
The standard comparison of theoretical calculation with
the phenomenological representation allows to extract the residue in
the senior pole and, therefore $B_B$. 
Eq.~(\ref{thfact}) and
Eq.~(\ref{thfull}) differ only due to non-factorizable corrections.

To find the nonfactorizable addition to $B_B$ from the sum rules 
we form ratios of the total and factorizable contributions.
On the ``theoretical'' side one finds
\be
\label{fintheory}
\frac{M_{th}(i,j)}{M_{th}^{f}(i,j)}
=1+\frac{\alpha_s}{4\pi}\frac{b^{nf}_{ij}}{1+\frac{\alpha_s}{4\pi}
b^{f}_{ij}}\, .
\ee
This ratio is mass-independent.
On the ``phenomenological'' side we have
\be
\frac{M_{ph}(i,j)}{M_{ph}^{f}(i,j)}=
\frac{B_B+R_B (z^j+z^i)+C_Bz^{i+j}}
{1+R^f(z^j+z^i)+C^f z^{i+j}}
\ee
where $z=m_B^2/(m_B^2+\Delta)$ is a parameter that describes the
suppression of higher state contributions. $\Delta$ is a gap 
between the squared masses of the $B$-meson and higher states. 
$R_B$, $C_B$, $R^f$ and $C^f$ 
are parameters of the model for higher state contributions within the sum
rule approach. 
In order to extract the non-factorizable
contribution to $B_B$ we write $B_B=1+\Delta B$.
Similarly, one can parameterize contributions to ``phenomenological'' moments 
due to higher $B$-meson states by writing $R_B=R^f+\Delta R$ and 
$C_B=C^f+\Delta C$. Clearly, $\Delta B=\Delta R=\Delta C=0$ in factorization. 
The final formula for the determination of $\Delta B$ reads
\[
\frac{\alpha_s}{4\pi} b^{nf}_{ij}
=\Delta B
+\Delta R (z^{j-2}+z^{i-2})
+\Delta C z^{i+j-4}
\]
where $\Delta R$ and $\Delta C$ are free parameters of the fit.
We take $\Delta=0.4 m_B^2$ for the $B$ meson 
two-point correlator. This corresponds to the duality interval of
$1~{\rm GeV}$ in energy scale for the analysis based on finite energy
sum rules~\cite{locduality}. The actual value of $\Delta B$
has been extracted using the least-square fit of all available
moments.
Estimating all uncertainties we
finally find the NLO non-factorizable QCD corrections to
$\Delta B$ due to perturbative contributions to the sum rules to be
\[
\Delta B=(6\pm 1)\frac{\alpha_s(m)}{4\pi} .
\]
We checked the stability of the sum rules which lead 
to a prediction of $\Delta B$.
For $m_b=4.8~\GeV$ and $\alpha_s(m_b)=0.2$~\cite{Penin:1999kx} 
one finds $\Delta B=0.1$.

\section{Conclusion}
To conclude, the $B^0 -\bar B^0$ mixing matrix element 
has been evaluated in the 
framework of QCD sum rules for three-point functions at NLO
in perturbative QCD. The effect of radiative corrections on $B_B$ is under 
complete control within pQCD and amounts to approximately $+10$\% of
the factorized value. 
The calculation can be further improved with the evaluation of 
higher moments. The result is sensitive to the parameter $z$ or to
the magnitude of the mass gap $\Delta$ used in the parametrization of the
spectrum.

\end{document}